\documentclass[twoside,twocolumn,9pt]{article}
\usepackage{extsizes}
\usepackage[super,sort&compress,comma]{natbib} 
\usepackage[version=3]{mhchem}
\usepackage[left=1.5cm, right=1.5cm, top=1.785cm, bottom=2.0cm]{geometry}
\usepackage{balance}
\usepackage{mathptmx}
\usepackage{sectsty}
\usepackage{graphicx} 
\usepackage{lastpage}
\usepackage[format=plain,justification=justified,singlelinecheck=false,font={stretch=1.125,small,sf},labelfont=bf,labelsep=space]{caption}
\usepackage{float}
\usepackage{fancyhdr}
\usepackage{fnpos}
\usepackage[english]{babel}
\addto{\captionsenglish}{%
  
}
\usepackage{array}
\usepackage{droidsans}
\usepackage{charter}
\usepackage[T1]{fontenc}
\usepackage[usenames,dvipsnames]{xcolor}
\usepackage{setspace}
\usepackage[compact]{titlesec}
\usepackage{hyperref}
\usepackage{epstopdf}%This line makes .eps figures into .pdf - please comment out if not required.

\definecolor{cream}{RGB}{222,217,201}

\usepackage{color}
\usepackage{amsmath,amssymb}
\usepackage{upgreek}
\usepackage{changes}

\AppendGraphicsExtensions{.tif}
\begin{document}

\pagestyle{fancy}
\thispagestyle{plain}
\fancypagestyle{plain}{
%%%HEADER%%%
\renewcommand{\headrulewidth}{0pt}
}
%%%END OF HEADER%%%

%%%PAGE SETUP - Please do not change any commands within this section%%%
\makeFNbottom
\makeatletter
\renewcommand\LARGE{\@setfontsize\LARGE{15pt}{17}}
\renewcommand\Large{\@setfontsize\Large{12pt}{14}}
\renewcommand\large{\@setfontsize\large{10pt}{12}}
\renewcommand\footnotesize{\@setfontsize\footnotesize{7pt}{10}}
\makeatother

\renewcommand{\thefootnote}{\fnsymbol{footnote}}
\renewcommand\footnoterule{\vspace*{1pt}% 
\color{cream}\hrule width 3.5in height 0.4pt \color{black}\vspace*{5pt}} 
\setcounter{secnumdepth}{5}

\makeatletter 
\renewcommand\@biblabel[1]{#1}            
\renewcommand\@makefntext[1]% 
{\noindent\makebox[0pt][r]{\@thefnmark\,}#1}
\makeatother 
\renewcommand{\figurename}{\small{Fig.}~}
\sectionfont{\sffamily\Large}
\subsectionfont{\normalsize}
\subsubsectionfont{\bf}
\setstretch{1.125} %In particular, please do not alter this line.
\setlength{\skip\footins}{0.8cm}
\setlength{\footnotesep}{0.25cm}
\setlength{\jot}{10pt}
\titlespacing*{\section}{0pt}{4pt}{4pt}
\titlespacing*{\subsection}{0pt}{15pt}{1pt}
%%%END OF PAGE SETUP%%%

%%%FOOTER%%%
\fancyfoot{}
\fancyfoot[LO,RE]{\vspace{-7.1pt}\includegraphics[height=9pt]{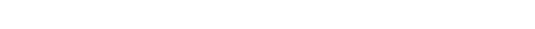}}
\fancyfoot[CO]{\vspace{-7.1pt}\hspace{11.9cm}\includegraphics{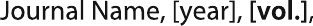}}
\fancyfoot[CE]{\vspace{-7.2pt}\hspace{-13.2cm}\includegraphics{head_foot/RF}}
\fancyfoot[RO]{\footnotesize{\sffamily{1--\pageref{LastPage} ~\textbar  \hspace{2pt}\thepage}}}
\fancyfoot[LE]{\footnotesize{\sffamily{\thepage~\textbar\hspace{4.65cm} 1--\pageref{LastPage}}}}
\fancyhead{}
\renewcommand{\headrulewidth}{0pt} 
\renewcommand{\footrulewidth}{0pt}
\setlength{\arrayrulewidth}{1pt}
\setlength{\columnsep}{6.5mm}
\setlength\bibsep{1pt}
%%%END OF FOOTER%%%

%%%FIGURE SETUP - please do not change any commands within this section%%%
\makeatletter 
\newlength{\figrulesep} 
\setlength{\figrulesep}{0.5\textfloatsep} 

\newcommand{\topfigrule}{\vspace*{-1pt}% 
\noindent{\color{cream}\rule[-\figrulesep]{\columnwidth}{1.5pt}} }

\newcommand{\botfigrule}{\vspace*{-2pt}% 
\noindent{\color{cream}\rule[\figrulesep]{\columnwidth}{1.5pt}} }

\newcommand{\dblfigrule}{\vspace*{-1pt}% 
\noindent{\color{cream}\rule[-\figrulesep]{\textwidth}{1.5pt}} }

\makeatother
%%%END OF FIGURE SETUP%%%

%%%TITLE, AUTHORS AND ABSTRACT%%%
\twocolumn[
  \begin{@twocolumnfalse}

\vspace{1em}
\sffamily
\begin{tabular}{m{4.5cm} p{13.5cm} }

\includegraphics{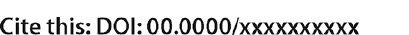} & \noindent\LARGE{\textbf{Interactions in protein solutions close to liquid-liquid phase separation: Ethanol reduces attractions via changes of the dielectric solution properties}} \\%Article title goes here instead of the text "This is the title"
\vspace{0.3cm} & \vspace{0.3cm} \\

 & \noindent\large{Jan Hansen,\textit{$^{a}$} 
 Rajeevann Uthayakumar,\textit{$^{a}$} 
Jan Skov Pedersen,\textit{$^{b}$} 
 Stefan U. Egelhaaf,\textit{$^{a}$} and
Florian Platten,$^{\ast}$\textit{$^{a,c}$}}  \\%Author names go here instead of "Full name", etc.

\includegraphics{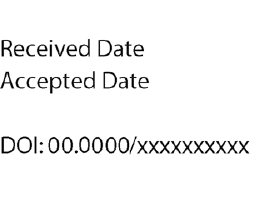} & \noindent\normalsize{Ethanol is a common protein crystallization agent, precipitant, and denaturant, but also alters the dielectric properties of solutions. 
While ethanol-induced unfolding is largely ascribed to its hydrophobic parts, its effect on protein phase separation and inter-protein interactions remains poorly understood.
Here, the effects of ethanol and NaCl on the phase behavior and interactions of protein solutions are studied in terms of the metastable liquid-liquid phase separation (LLPS) and the second virial coefficient $B_2$ using lysozyme solutions. 
Determination of the phase diagrams shows that the cloud-point temperatures are reduced and raised by the addition of ethanol and salt, respectively. 
The observed trends can be explained using the extended law of corresponding states as changes of $B_2$. 
The results for $B_2$ agree quantitatively with those of static light scattering and small-angle X-ray scattering experiments. 
Furthermore,  $B_2$ values calculated based on inter-protein interactions described by the Derjaguin--Landau--Verwey--Overbeek (DLVO) potential and considering the dielectric solution properties and electrostatic screening due to the ethanol and salt content quantitatively agree with the experimentally observed $B_2$ values.
} \\%The abstrast goes here instead of the text "The abstract should be..."
\end{tabular}
 \end{@twocolumnfalse} \vspace{0.6cm}  ]
%%%END OF TITLE, AUTHORS AND ABSTRACT%%%

%%%FONT SETUP - please do not change any commands within this section
\renewcommand*\rmdefault{bch}\normalfont\upshape
\rmfamily
\section*{}
\vspace{-1cm}

%%%FOOTNOTES%%%

\footnotetext{\textit{$^{a}$~Condensed Matter Physics Laboratory, Heinrich Heine University, Universit\"atsstraße~1, 40225 D\"usseldorf, Germany; E-mail: florian.platten@hhu.de};}
\footnotetext{\textit{$^{b}$~iNANO Interdisciplinary Nanoscience Center and Department of Chemistry, Aarhus University, DK-8000 Aarhus C, Denmark;}}
\footnotetext{\textit{$^{c}$~Institute of Biological Information Processing (IBI-4: Biomacromolecular Systems and Processes), Forschungszentrum J\"ulich, Wilhelm-Johnen-Straße, 52428 J\"ulich, Germany. }}

%Please use \dag to cite the ESI in the main text of the article.
%If you article does not have ESI please remove the the \dag symbol from the title and the footnotetext below.
%\footnotetext{\Q{\dag~Electronic Supplementary Information (ESI) available: Analysis of the crystallization entropy. See DOI: 10.1039/cXCP00000x/}}
%additional addresses can be cited as above using the lower-case letters, c, d, e... If all authors are from the same address, no letter is required

%\footnotetext{\ddag~Additional footnotes to the title and authors can be included \textit{e.g.}\ `Present address:' or `These authors contributed equally to this work' as above using the symbols: \ddag, \textsection, and \P. Please place the appropriate symbol next to the author's name and include a \texttt{\textbackslash footnotetext} entry in the the correct place in the list.}

%%%END OF FOOTNOTES%%%

%%%MAIN TEXT%%%%

\section{Introduction}
 
When the attractions between protein molecules are strong enough, protein solutions can undergo liquid-liquid phase separation (LLPS) into two coexisting phases, one enriched and one depleted in proteins. 
Such protein phase separation has severe implications in fundamental and applied fields of research, including cell biology, medicine, pharmaceutical industry, food processing, and protein crystallography. 
For example, LLPS is exploited in vivo: Subcellular compartments, so-called membraneless organelles, are formed via LLPS in the cytosol, representing a way of intracellular organization and regulation of biochemical reactions.\cite{Hyman2014,Brangwynne2015}
Furthermore, genetic mutations or altered physicochemical conditions inside a cell are likely to affect inter-protein interactions and thus to disturb LLPS.\cite{Shin2017,Alberti2021}
However, LLPS can also modulate the pathways and kinetics of pathological protein aggregation leading to severe conditions for the patients.\cite{Babinchak2020,Ray2020}
Protein solutions exhibit LLPS not only in vivo, but also in vitro.\cite{Ishimoto1977,Muschol1997}
For example, antibodies, which are used as biopharmaceuticals in the treatment of various diseases,\cite{Nelson2010} can undergo LLPS due to nonspecific antibody--antibody interactions.\cite{Wang2013,Wang2014,DaVela2017} 
The dense-phase LLPS droplets might impair specific antibody-receptor interactions, enhance solution viscosity and cause immungenicity, posing a major challenge to the formulation development.\cite{Salinas2010,Raut2016}
LLPS can also be employed for identifying conditions under which high-quality crystals grow, which are needed for crystallographic structure determination.\cite{Chayen2008} Yet, attempts to crystallize proteins are still frequently based on trial and error.
Close to the LLPS binodal enhanced protein crystal nucleation rates have been predicted by simulations and experiments,\cite{Wolde1997,Galkin2000}
and therefore the location of the LLPS boundary has been regarded as a predictor for optimized crystallization conditions,\cite{Vliegenthart2000,Zhang2012b}
as neither too weak nor too strong attractions are considered to be well suited for crystallization.\cite{George1994} 
Within the LLPS binodal, liquid-droplet nucleation or spinodal decomposition are typically faster than crystallization, thus leading to a two-step crystallization mechanism\cite{Vekilov2010a,Zhang2012a} and  altering crystallization kinetics\cite{Liu2010,Sauter2015}.

Net attractions sufficient to induce LLPS can be achieved or avoided by dedicated changes of the physicochemical properties of the solution, e.g., by adding salts,\cite{Broide1996,Sedgwick2005} excipients,\cite{Raut2016a} or non-aqueous solvents\cite{Goegelein2012,Platten2015b}.
Moreover, organic solvents, such as alcohols, can act as precipitants\cite{Timasheff1988,Yoshikawa2012} and as crystallization agents\cite{Cohn1947,McPherson2004}. 
For example, in  blood plasma fractionation,\cite{Cohn1946,Green1955} moderate ethanol concentrations (up to 40~vol.\%) are used to obtain therapeutic protein products.
If added at high concentrations or used at elevated temperatures, ethanol can destabilize and unfold proteins,\cite{Brandts1967,Parodi1973} which might even lead to amyloid fibril formation.\cite{Goda2000,Holley2008,Giugliarelli2015}
Its effects on individual protein molecules\cite{Ikeda1970,Thomas1993} are largely ascribed to its hydrophobic properties.\cite{Nozaki1971} 
Ethanol is composed of a hydrophobic ethyl group and a hydrophilic hydroxyl group and can thus interact favorably with non-polar groups.\cite{Yoshikawa2012} 
However, far less (mechanistic) insight is established concerning the effects on inter-protein interactions, as relevant, e.g., for LLPS.

The second virial coefficient $B_2$ represents an integral measure of the inter-protein interactions, which for a spherosymmetric potential $U(r)$ with center-to-center distance $r$ reads
\begin{equation}\label{eq:b}
B_2 = 2 \pi \int_0^\infty \left( 1 - \exp{\left[-\frac{U(r)}{k_\text{B} T} \right]} \right) r^2 \text{d}r \, 
\end{equation}
with thermal energy $k_\text{B}T$.
Upon addition of moderate alcohol concentrations,
Liu et al.\cite{Liu2004} observed an initial increase and then a plateau of $B_2$, whereas Kundu et al.\cite{Kundu2017} reported a decrease for similar solution conditions. 
Direct and indirect mechanisms could be responsible for such effects: 
Alcohol molecules might act as protein-binding ligands or induce nonlocal changes of the dielectric solvent properties. 
In view of the inconsistent results,\cite{Liu2004,Kundu2017}
both the effect and the underlying mechanism remain controversial.

Concepts developed in soft-matter physics\cite{Gunton2007,McManus2016,Stradner2020} have proven helpful to rationalize the inter-protein interactions and the related phase behavior.
Experimental protein phase diagrams, including LLPS phase coexistence curves (binodals), are strikingly similar to those of colloids with short-ranged attractions,\cite{Asherie1996,Poon1997} as encountered, e.g., in square-well (SW) fluids\cite{Asherie1996,Lomakin1996,Duda2009,Platten2015} or patchy particle systems.\cite{Sear1999,Lomakin1999,Liu2007,Goegelein2008,Kastelic2015} 
Furthermore, the structure factor of protein solutions close to phase separation has been described by Baxter's sticky particle model,\cite{Zhang2012b,Moeller2014,Wolf2014} for which an approximate analytical description is available.\cite{Regnaut1989,Menon1991}
For colloids with short-ranged attractions, an extended law of corresponding states (ELCS) has been suggested by Noro and Frenkel,\cite{Noro2000} according to which short-ranged attractive systems can be mapped onto an equivalent SW system, and the applicability of the ELCS to the binodals of protein solutions has been demonstrated.\cite{Platten2015}
The ELCS mapping thus reflects the insensitivity to the specific shape of the coarse-grained model potential, and in particular, it allows the estimation of $B_2$ based on cloud-point measurements.\cite{Platten2016}
In this context, the Derjaguin--Landau--Verwey--Overbeek (DLVO) theory has helped to rationalize the dependence of inter-protein interactions on simple salts, solvents or $p$H.\cite{Muschol1995,Poon2000,Sedgwick2007,Pellicane2012,Kumar2019}

In the present work, the effects of moderate ethanol concentrations on protein molecules, inter-protein interactions as well as LLPS coexistence curves are studied using lysozyme in brine as a model system. 
Small-angle X-ray scattering (SAXS) was used for determining the form factor of protein molecules, thus confirming that the alcohol and salt have no influence on the structure of the individual molecules on the relevant length scales.
Cloud-point measurements are used to locate the LLPS binodal and to estimate $B_2$ exploiting the ELCS.
SAXS is also used to study the structure factor of concentrated protein solutions close to phase separation.
From the analysis of the SAXS data, $B_2$ is determined as a function of the ethanol content. 
The results confirm a universal temperature dependence of $B_2$ with respect to the critical LLPS temperature, as suggested by the ELCS.
The dependence of $B_2$ on ethanol and salt content is quantitatively described by the DLVO theory taking only changes of the dielectric solution properties and the salt concentration into account.
This work thus aims at a consistent picture of protein phase separation, a mechanistic explanation of solvent effects on inter-protein interactions and a resolution of controversial previous results.

%\newpage
\section{Experimental methods}

\subsection{Sample preparation}

Hen egg-white lysozyme was purchased from Sigma-Aldrich (prod. no. L6876)  and used without further purification.
For few SAXS experiments, lysozyme purchased from Roche Diagnostics (prod. no. {10837059001}) was used, which led to consistent findings.
Sodium chloride (NaCl), sodium acetate (NaAc) and ethanol (EtOH) were of reagent grade quality and used as received. 
Ultrapure water with a minimum resistivity of $18~\text{M}\Omega\,\text{cm}$ was prepared using a water purification system.  

Water--ethanol mixtures containing 50~mM NaAc were used as buffer solutions and adjusted to $p$H reading 4.5 by adding small amounts of hydrochlorid acid. 
At $p$H 4.5 each lysozyme molecule carries approximately 11.4 positive net charges.\cite{Tanford1972}
Concentrated protein stock solutions were prepared by ultrafiltration, as described previously.\cite{Platten2015b}
The protein, ethanol and salt content of the stock solutions was checked by refractometry.
With respect to $p$H value (4.5) and NaCl concentrations (0.7~M and 0.9~M), solution conditions are chosen to resemble those of our previous studies\cite{Hansen2016,Hentschel2021} to allow for a quantitative comparison.
Concerning the ethanol content, low and moderate concentrations (up to $30~\text{vol.}\%$ in increments of $10~\text{vol.}\%$) are considered, similar to those of Liu et al.\cite{Liu2004} and Kundu et al.\cite{Kundu2017}
At $p$H 2.2, where lysozyme is expected to be less stable than at $p$H 4.5, ethanol-induced (partial) unfolding of lysozyme was only observed for ethanol concentrations larger than $30~\text{vol.}\%$.\cite{Sasahara2006}
Samples were prepared by mixing appropriate amounts of lysozyme, buffer and NaCl stock solutions. 
Protein concentrations $c_\text{p}$ are related to the protein volume fraction $\phi = c_\text{p}/\rho_\text{p}$, where $\rho_\text{p}^{-1} = 0.740~\text{cm}^3/\text{g}$ is the specific volume of lysozyme.\cite{Platten2015b} 
Mixing was performed at a temperature above the solution cloud-points to prevent immediate phase separation, typically at room temperature $({21 \pm 2})~^\circ$C.
Due to the high salt content, the samples were prone to crystallization\cite{Sedgwick2005} and hence investigated immediately after preparation.
Cloud points were typically studied using three independently prepared samples for each condition in order to allow for a statistical analysis.
For SAXS, some of the samples were measured more than once in order to check the reproducibility of our results.

\subsection{Cloud-point temperature measurements}
Metastable LLPS coexistence curves were determined by cloud-point temperature measurements. 
Samples with a typical volume of 0.1~mL were filled into thoroughly cleaned glass capillary tubes, sealed, and placed into a thermostated water bath at a temperature well above the cloud-point. 
A wire thermometer was mounted in a separate, but closely placed glass tube filled with 0.1~mL water. 
Then, the temperature of the water bath was gradually lowered and the sample solution visually observed. 
The cloud point was identified by the sample becoming turbid. 
Further details have been given previously.\cite{Platten2015b}

\subsection{Small-angle X-ray scattering: Instrumentation}
Small-angle X-ray scattering (SAXS) was employed to determine the form factor of individual protein molecules to reveal possible shape or size changes as well as the structure factor characterizing the effective inter-protein interactions at temperatures close to, but above the solution cloud-points.  
SAXS experiments were performed using the laboratory-based facilities at the Interdisciplinary Nanoscience Center (iNANO) at Aarhus University, Denmark,\cite{Behrens2014} as well as at Center for Structural Studies at Heinrich Heine University Düsseldorf, Germany.
In Aarhus, a NanoSTAR SAXS camera (Bruker AXS) optimized for solution scattering\cite{Pedersen2004} with a home-built scatterless pinhole in front of the sample\cite{Lyngso2021} was used to measure the scattered intensity of sample and buffer solutions.
The solutions were filled in a thin flow-through glass capillary and thermostated using a Peltier element (Anton Paar). 
In Düsseldorf, SAXS measurements on sample and buffer solutions were performed on a XENOCS 2.0 device with a Pilatus 3 300K detector.
The solutions were injected into a thin flow-through capillary cell mounted on a thermal stage.
Experiments were performed at $20.0~^\circ\text{C}$ and $25.0~^\circ\text{C}$, respectively. 
Typical acquistion times of 10 and 5~min were used for dilute and concentrated solutions (typically, $6$ and $70~\text{mg/mL}$), respectively.
The data were background subtracted and converted to absolute scale using water in Aarhus\cite{Pedersen2004} and glassy carbon in Düsseldorf as standards.
The final intensity is displayed as a function of the magnitude of the scattering vector, $Q = \frac{4\pi}{\lambda_0}\sin(\theta)$, where the X-ray wavelength, $\lambda_0$, is $1.54~\text{\AA}$ and $2\,\theta$ is the angle between the incident and scattered X-rays and calibration was performed using silver behenate.

\subsection{Small-angle X-ray scattering: Data analysis}
Protein molecules tend to have anisotropic shapes; according to X-ray crystallography, lysozyme is approximately a prolate ellipsoid with an extension of $30 \times 30 \times 45~\text{\r{A}}^3$.\cite{Blake1965}
For a monodisperse solution of particles with only a small anisotropy, the interactions can be assumed to be independent of the orientation.
Then, the absolute scattered intensity $I(Q)$ can be described by the decoupling approximation:\cite{Kotlarchyk1983,Chen1986,Pedersen1997} 		
	\begin{equation}\label{eq:1}
	I(Q) = K\, c_\text{p}\, M\, P(Q) \, S_\text{eff}(Q) \, .
	\end{equation}
The $Q$ dependence of the scattered intensity is due to intra-particle and inter-particle interference effects quantified by $P(Q)$ and $S(Q)$, respectively. 
The form factor $P(Q)= \langle A^2(Q) \rangle_{\Omega}$ is obtained from the form factor amplitude $A(Q)$ averaged (as denoted by brackets) over particle orientations $\Omega$, and the effective structure factor reads
\begin{equation}
S_\text{eff}(Q)=1+ \frac{ \langle A(Q) \rangle^2_{\Omega} }{ \langle A^2(Q) \rangle_{\Omega} }  \left[ S(Q) - 1 \right] \, ,
\end{equation}
where $S(Q)$ is the structure factor of an effective one-component system.
The magnitude of the absolute scattered intensity depends on the particle (protein) mass concentration $c_\text{p}$, its molecular weight $M=14\,320~\text{g/mol}$, and the
contrast factor $K \sim (\Delta \rho)^2$ related to the electron density difference $\Delta \rho$ between particle and solvent, which can be computed\cite{Whitten2008,Sarachan2013}.

For very dilute systems, $S(Q) \approx 1$ and the $Q$ dependence of $I(Q)$ is determined by the size, shape and structure of the individual particles via $P(Q)$.
In particular, the radius of gyration $R_\text{g}$, a measure of the particle size, can be inferred from the low-$Q$ scattering.
To describe the shape and structure of the lysozyme molecules, two different models for $P(Q)$ are considered here.
On a coarse level,\cite{Pedersen1997} the form factor of lysozyme can be modelled as a prolate ellipsoid of revolution with minor and major axes as parameters. 
Since the atomic coordinates of lysozyme are known (PDB file {1LYZ} \cite{Diamond1974}), the form factor can be calculated accurately using the programme CRYSOL,\cite{Svergun1995} which calculates the excess scattering and adds a $3~\text{\r{A}}$ hydration shell with the shell electron density $\rho_\text{sh}$ as a parameter.

In concentrated solutions, the structure factor $S(Q)$ contains information on the spatial arrangement of the particles and thus reflects inter-particle interactions.
In a one-component system, the coexistence of two stable or metastable fluid phases is only possible if the particle interactions are net attractive. 
The square-well (SW) potential arguably represents the simplest model to describe the effective interactions of such a system.\cite{Asherie1996} 
It consists of a hard-core repulsion of range $\sigma$ (the diameter of the particle), which leads to excluded volume effects, and a constant attractive part, which has depth $\epsilon$ and extends to a distance $\lambda \sigma$ from the center.
The adhesive hard-sphere (AHS) potential proposed by Baxter represents a specific limit of the SW potential:\cite{Baxter1968} the SW depth $\epsilon$ becomes infinite while the SW width $(\lambda-1)\sigma$ becomes infinitesimal (i.e., $\lambda \rightarrow 1$), such that the contribution to $B_2$ remains finite and nonzero:
\begin{equation}
\frac{U_\text{AHS}(r)}{k_\text{B}T} = \lim_{\lambda\rightarrow 1}
 \begin{cases}
      \infty    & \text{for} \quad 0<r<\sigma \, ,\\
      \ln[ 12 \tau (1-1/\lambda)]  & \text{for} \quad \sigma<r<\lambda\sigma \, , \\
      0         & \text{for} \quad r>\lambda\sigma \, .
   \end{cases} 
\end{equation}
The corresponding Boltzmann factor reads
\begin{equation}
\exp[-U_\text{AHS}(r) / k_\text{B}T] = \Theta(r-\sigma) + \frac{\sigma}{12\tau} \delta(r-\sigma) \, ,
\end{equation}
where the Heaviside step function $\Theta$ takes into account that overlap of the  hard cores is not possible and the Dirac delta function favors contacts due to the stickiness parameter $\tau$.
The net interactions of the AHS potential, as quantified by  $B_2$ (Eq.~\ref{eq:b}), are determined by only one parameter: 
\begin{equation}\label{eq:stick}
b_2 = 1-\frac{1}{4\tau} \,,
\end{equation}	
where $b_2=B_2/B_2^\text{HS}$ is the second virial coefficient of the system $B_2$ normalized by that of the corresponding hard-sphere system $B_2^\text{HS}=(2\pi/3)\,\sigma^3$.
Approximate analytical descriptions of the structure factor of adhesive hard spheres in the Percus-Yevick closure are available\cite{Regnaut1989,Menon1991,Menon1991b} and commonly used to model scattering data of short-range attractive systems\cite{Bergstrom1999,Eberle2012}.	
The AHS structure factor depends on the effective particle diameter $\sigma$, the stickiness $\tau$ and the particle volume fraction $\phi$.
The effective particle diameter $\sigma$ is identified with the diameter of a sphere that has the same volume as the ellipsoid determined by form-factor modelling in a dilute solution.
The volume fraction $\phi$ is related to the protein concentration $c_\text{p}$.
Thus, only one fitting parameter, $\tau$, has to be determined.
To model the structure factor of proteins, inter-protein interactions are often described by the sum of an attractive and a repulsive hard-core Yukawa potential each with its own range and interaction strength parameter.\cite{Tardieu1999,Liu2005c,Chen2007,Kundu2016}
However, close to LLPS, the interactions are dominated by net attractions and, hence, the simpler AHS description is favored here, allowing for a direct determination of $b_2$, 

The scattered intensity based on Eq.~(\ref{eq:1}) with a constant scattering background is fitted to the measured scattered intensity using a least-square routine.
Since background subtraction is particularly delicate at very low $Q$, model fits are compared with experimental data for $Q \geq 0.015~\text{\AA}^{-1}$.
The contrast factor of lysozyme in the different water--ethanol mixtures is calculated using the MULCh software.\cite{Whitten2008}
To account for the experimental uncertainty in $c_\text{p}$, a deviation of up to $10~\%$ from its nominal value is allowed in fitting.

%\newpage
\section{Results and Discussion}
 
First, the effect of moderate ethanol concentration on the size and shape  of individual protein molecules is investigated by SAXS.
In a second step, LLPS coexistence curves are determined for various ethanol and salt compositions.
Then, the structure factor of moderately concentrated systems close to LLPS is examined by SAXS measurements of concentrated samples. 
Both from the cloud-point measurements (by exploiting the ELCS) and from the SAXS data (by applying Baxter's AHS model), the normalized second virial coefficient $b_2$ is inferred. 
Finally, the dependence of $b_2$ on ethanol and salt content is rationalized based on DLVO theory.

\subsection{Lysozyme molecular structure in water--ethanol mixtures}\label{sec:form}

The scope of the present work is on the LLPS of folded, globular proteins,\cite{Vekilov2010} whose interactions are tuned by the addition of ethanol and NaCl.
However, ethanol can also denature and aggregate proteins.\cite{Parodi1973,Tanaka2001,Nemzer2013}  
Therefore, the shape and size of individual lysozyme molecules in water--ethanol mixtures are determined.

\begin{figure}[h!]
\includegraphics[width=\columnwidth]{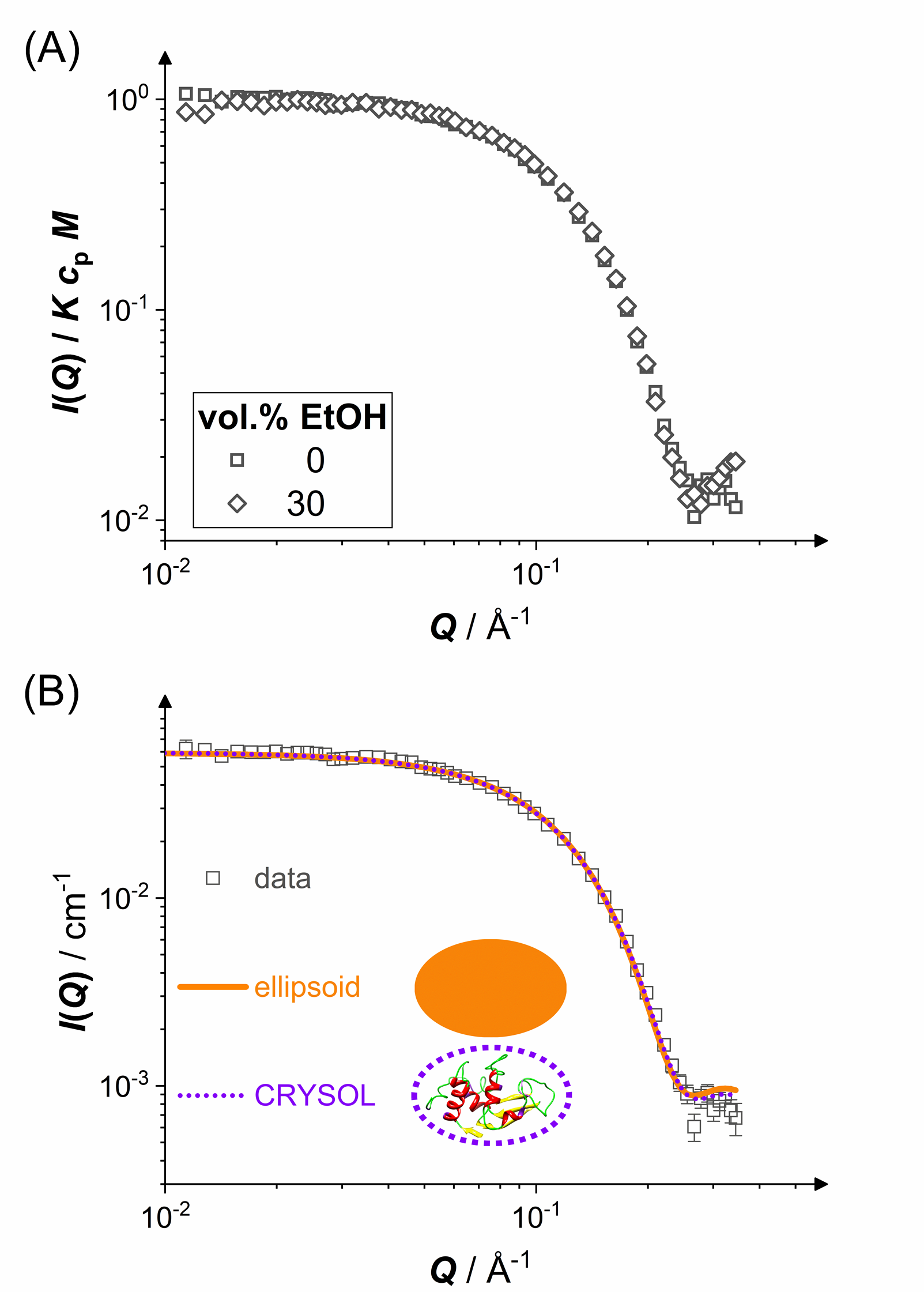} 
  \caption{Form factor of lysozyme molecules ($c_\text{p} \approx 6~\text{mg/mL}$) in brine (0.9 M NaCl) without and with $30~\text{vol.}\%$ EtOH (squares and diamonds, respectively): (A) Scattered intensity $I(Q)$ normalized by the contrast factor $K$, the molecular weight $M$, and the protein concentration $c_\text{p}$ as a function of the magnitude of the scattering vector $Q$.  (B) Scattered intensity $I(Q)$ as experimentally determined (squares as in (A)) and model fits (lines as indicated). Schematic drawings (not to scale) illustrate the two models. For the crystal structure, only the backbone is shown.}
  \label{fig:form}
\end{figure}

Figure~\ref{fig:form} shows the scattered intensity $I(Q)$ of dilute protein solutions ($c_\text{p}\approx 6~\text{mg/mL}$) under two particular conditions: (i) proteins in an aqueous solution with 0.9~M NaCl, which is used to screen electrostatic repulsions (squares), and (ii) proteins in a water--ethanol mixture with the highest ethanol concentration used in this work ($30~\text{vol.}\%$ EtOH) with 0.9~M NaCl (diamonds).
Experiments on intermediate ethanol concentrations show similar behavior.
In Fig.~\ref{fig:form}(A), the experimental data for both conditions are shown with $I(Q)$ normalized by the ethanol-dependent contrast factor $K$, the molecular weight $M$ and the protein concentration $c_\text{p}$.
Since $S(Q)\approx 1$ in this case, Eq.~(\ref{eq:1}) implies that the data reflect the form factor $P(Q)$. 
The data obtained for the two different conditions do not show any significant difference in the covered $Q$ range, indicating that the shape and size of the lysozyme molecules are not affected by $30~\text{vol.}\%$ ethanol. 
The $Q$ dependence of the data exhibts a plateau with $I(Q)/K\,c_\text{p}\,M\approx P(Q)\approx 1$ at low $Q$ and a minimum at high $Q$ which suggests a globular object.
A Guinier analysis indicates radii of gyration $R_\text{g} = 15.5~\text{\AA}$ ($0~\text{vol.}\%$ EtOH) and $R_\text{g} = 14.9~\text{\AA}$ ($30~\text{vol.}\%$ EtOH).
Radii of gyration determined in repeat measurements and experiments at intermediate ethanol concentrations range from approximately $15$ to $17~\text{\AA}$, indicating an experimental uncertainty of $1.1~\text{\AA}$. 

In Fig.~\ref{fig:form}(B), fits based on two different form factor models (lines) to the experimental data of the aqueous solution (symbols) are shown. 
The first form factor model is based on a prolate ellipsoid of revolution with a semi-minor axis $16.0~\text{\AA}$ and an axial ratio fixed at $1.5$ (solid line), which describes the experimental data reasonably well.
The second one is calculated via the CRYSOL programme\cite{Svergun1995} (dotted line) with $\rho_\text{sh}$ in agreement with previous work.\cite{Svergun1995} 
It quantitatively reproduces the experimental data and thus agrees with the ellipsoid model except for minor differences at very high $Q$.

The experimental data and the analysis demonstrate that lysozyme molecules retain their compact native shape under all conditions studied.
This is in line with a previous finding\cite{Sasahara2006} that higher ethanol concentrations are required to induce unfolding, which might change protein size and shape.

\subsection{Liquid-liquid phase separation of lysozyme solutions in water--ethanol mixtures}\label{sec:llps}

Samples might show a macroscopic phase transition accompanied by a clouding of the system which indicates LLPS.
The temperature at which the system becomes cloudy depends on the strength of the net attractions.
Higher cloud-point temperatures indicate stronger net attractions.
Cloud-point temperature measurements thus represent a simple way to characterize the inter-particle interactions.\cite{Platten2016}

\begin{figure}
\includegraphics[width=\columnwidth]{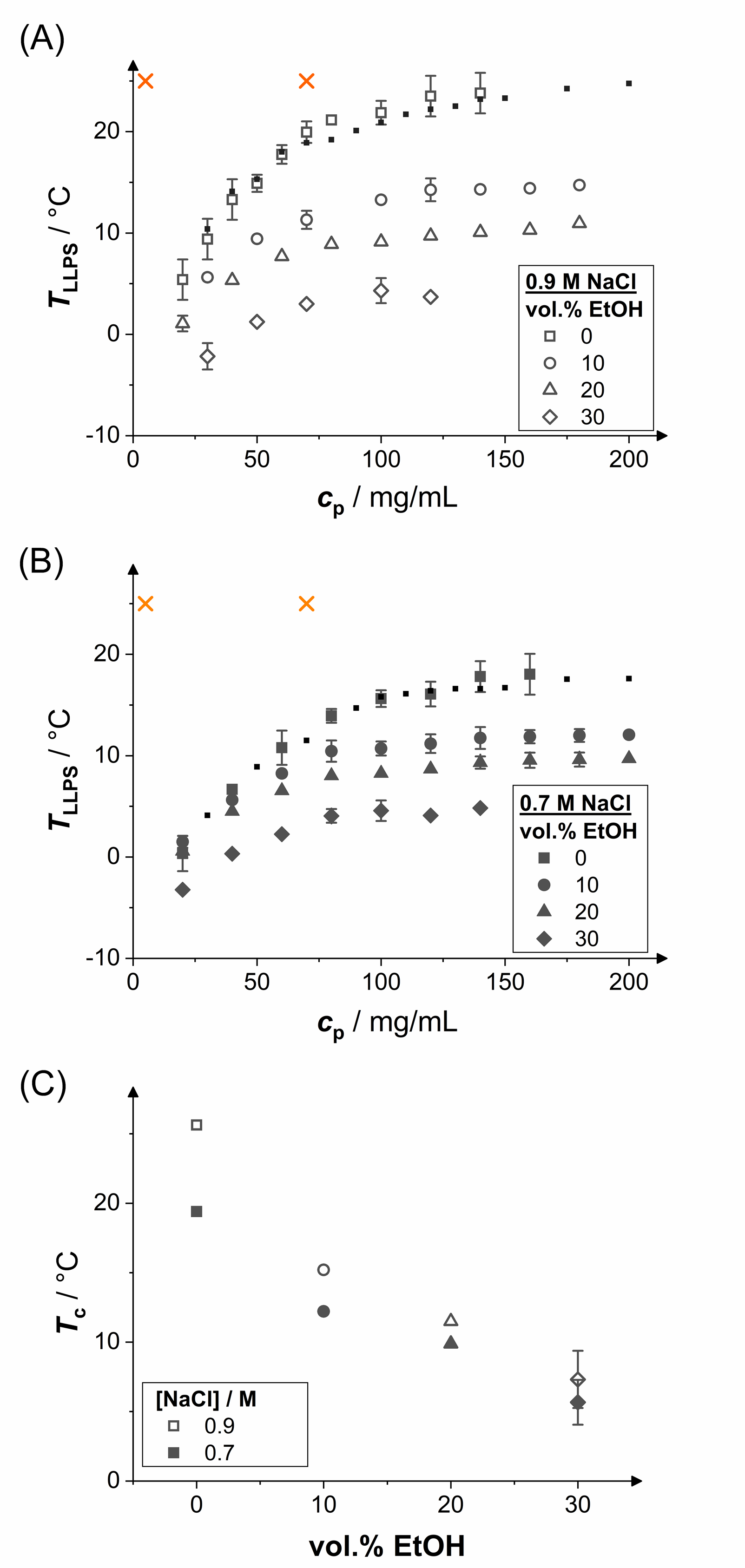} 
  \caption{Effect of ethanol on the LLPS. 
  Cloud-point temperature $T_\text{LLPS}$ as a function of protein concentration $c_\text{p}$ representing the LLPS coexistence curves in the presence of (A) 0.9~M NaCl and (B) 0.7~M NaCl as well as various ethanol concentrations as indicated (large symbols). 
  Literature data\cite{Goegelein2012} in the absence of ethanol (small symbols). 
  Typical solution conditions probed in SAXS experiments (crosses); at this temperature, $b_2$ is also estimated based on the cloud-points.  
  (C) Estimated critical temperature $T_\text{c}$ as a function of the ethanol content for the two NaCl concentrations as indicated.}
  \label{fig:state}
\end{figure}

Figure~\ref{fig:state}(A) and (B) show the low-volume fraction branch of LLPS phase coexistence curves of lysozyme solutions in brine (0.9 and 0.7~M NaCl represented by open and closed large symbols, respectively) with and without ethanol being added.
In the absence of ethanol, the data agree with literature results (small symbols).\cite{Goegelein2012}
With increasing protein concentration $c_\text{p}$, the cloud-point temperature $T_\text{LLPS}$ first increases steeply, reflecting an enhanced effect of inter-protein attractions.
Then, $T_\text{LLPS}$ saturates at high protein concentrations, indicating the proximity to the critical point.\cite{Platten2015}
In the latter case, critical scalings can be used for describing the $T-c_\text{p}$ dependence:\cite{Stanley1971,Broide1996,Muschol1997,Hansen2016}
\begin{equation} \label{eq:tc}
\left| \frac{c_\text{c}-c_\text{p}}{c_\text{c}} \right| = a \left(\frac{T_\text{c}-T_\text{LLPS}}{T_\text{c}}\right)^\beta 
\end{equation} 
with the critical temperature $T_\text{c}$, the critical concentration $c_\text{c}$, the critical exponent for binary demixing $\beta$ and a fitting parameter $a$.  
By fitting Eq.~(\ref{eq:tc}) to the data, $T_\text{c}$ can be estimated. 
To minimize the number of free parameters, $\beta$ is set to its renormalization group-theory value ($\beta=0.325$) and $c_\text{c}$ is fixed to a previously determined value ($c_\text{c}=270~\text{mg/mL}$)\cite{Platten2015}, as changes of $c_\text{c}$ with ethanol concentration are expected to be small.
To avoid distortions due to the off-critical part of the binodal, cloud-points at $c_\text{p} < 70~\text{mg/mL}$ have been excluded from the fit.
The resulting $T_\text{c}$ values are displayed in Fig.~\ref{fig:state}(C) as a function of the ethanol content for the two different NaCl concentrations.
Upon addition of ethanol, $T_\text{c}$ decreases as does $T_\text{LLPS}$ in general (Fig.~\ref{fig:state}(A,B)).
This indicates that ethanol reduces the net attractions between lysozyme molecules.

The attractions can be quantified by interaction parameters, such as the second virial coefficient $b_2$.
Based on cloud-point measurements, $b_2$ at a given temperature close to the binodal can be estimated by a comparison of the experimental binodal with those of short-ranged SW fluids,\cite{Platten2015}
as suggested by the extended law of corresponding states.\cite{Noro2000}
Following this approach, $b_2$ has been determined at the temperature indicated by crosses in Fig.~\ref{fig:state}(A,B) for the different solution compositions probed.
The results will be discussed in Section~\ref{sec:dlvo}.

\subsection{The interactions in protein solutions close to phase separation}\label{sec:saxs}

Pair interactions in protein solutions can be inferred from the concentration-dependence of the scattered intensity\cite{Bonnete1999} or from the $Q$ dependence of the scattering intensity through structure factor models\cite{Zhang2007}. 
However, close to the LLPS spinodal, critical or off-critical scattering contributions\cite{Bucciarelli2016} are expected to occur and analytical mean-field models usually employed to analyze scattering data are expected to fail. 
Thus, to be able to determine the pair interaction parameters through structure factor modelling, we focus on moderately concentrated solutions at temperatures that are at least a few degrees above the expected LLPS spinodal temperatures\cite{Manno2003}.

\begin{figure}
\includegraphics[width=\columnwidth]{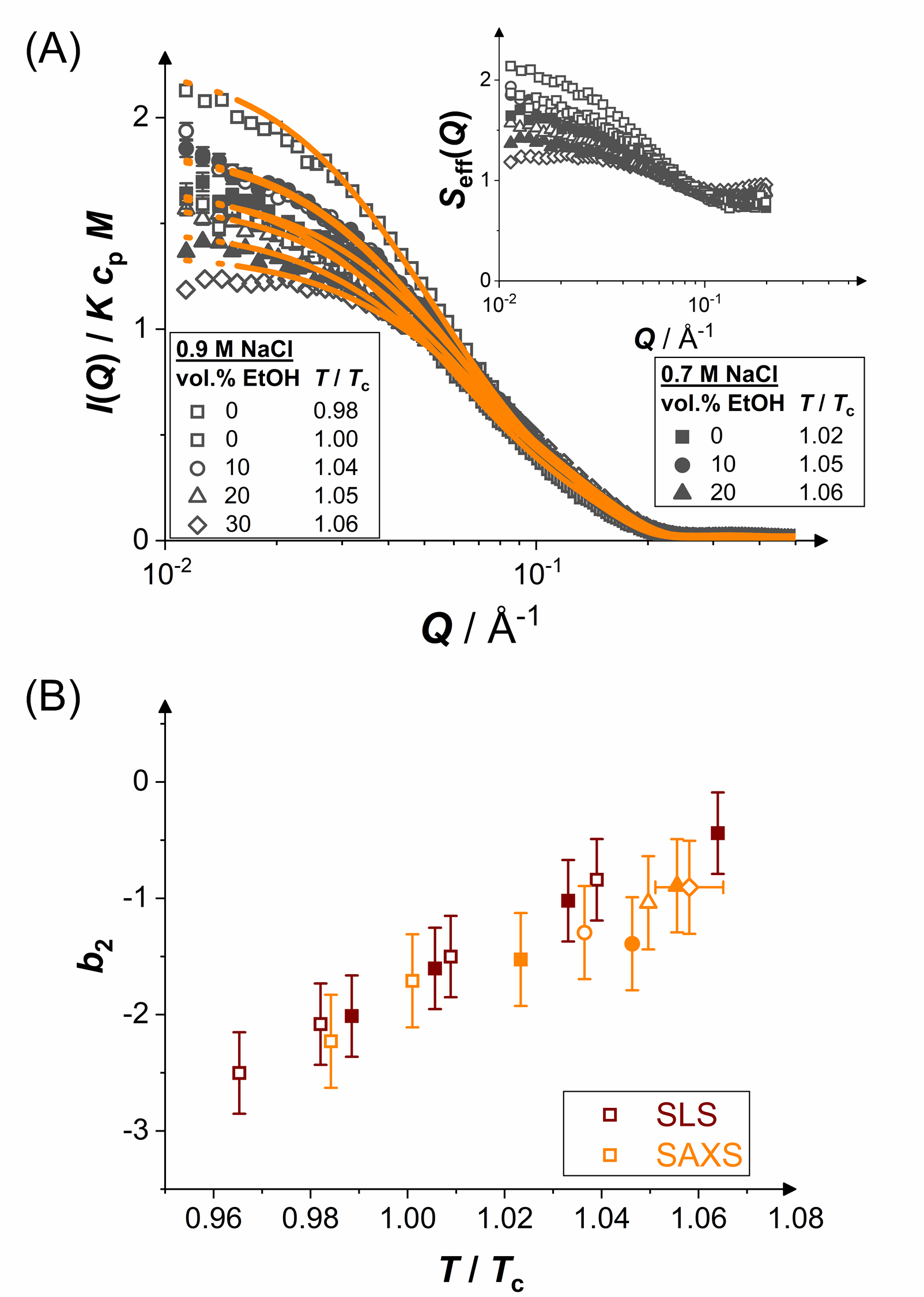} 
%includegraphics[width=\columnwidth]{saxs.tif}
%\includegraphics[width=\columnwidth]{b2Tc.tif}
  \caption{(A) Scattering vector-dependent normalized scattered intensity, $I(Q)/K\,c_\text{p}\,M$, of concentrated protein solutions ($c_\text{p} \approx 70~\text{mg/mL}$) close to phase separation (temperature $T$ relative to the respective critical temperature $T_\text{c}$, ethanol and salt content as indicated): experimental data (symbols) and model fits (lines). Inset: Effective structure factor $S_\text{eff}(Q)$ as inferred from the data, according to Eq.~(\ref{eq:1}). Only data  with $Q \leq 0.2~\mathrm{\AA}$ are shown, as they are very noisy beyond this value.
  (B) Normalized second virial coefficient $b_2$ as a function of temperature $T$ normalized by the critical temperature $T_\text{c}$. Data based on SAXS (orange symbols) and static light scattering\cite{Goegelein2012} (SLS, red symbols). 
  Open and closed symbols correspond to 0.9 and 0.7~M NaCl, respectively.
  Ethanol content is reflected in the symbol shape as in (A).}
  \label{fig:saxs}
\end{figure}

Figure~\ref{fig:saxs}(A) shows the normalized scattered intensity $I(Q)/K\,c_\text{p}\,M$ of concentrated protein solutions ($c_\text{p}\approx 70~\text{mg/mL}$ in the presence of ethanol and $50~\text{mg/mL}$ in the absence of ethanol) under conditions close to phase separation, as indicated in Fig.~\ref{fig:state}(A,B).
The experiments were performed at a fixed temperature.
However, due to the different ethanol and NaCl concentrations (as indicated by the symbol type and filling, respectively) and hence different $T/T_\text{c}$, the distance to the LLPS boundary increases as $T_\text{c}$ decreases with ethanol content (Fig.~\ref{fig:state}(C)) and decreases as $T_\text{c}$ increases with NaCl concentration.
In order to compare the different solution compositions with each other, all data are shown in a single graph.

According to Eq.~(\ref{eq:1}), the $Q$ dependence of the scattering curves reflects both the form and structure factor. 
Since $P(Q)$ was found to be unaltered by the different solution conditions, the variations in $I(Q)$ are largely due to changes of $S(Q)$.
At intermediate and high $Q$, the curves do not reveal marked differences.
However, at low $Q$, the scattered intensity tends to increase as $T/T_\text{c}$ decreases and hence the distance to the LLPS boundary also decreases.
The effective structure factors $S_\text{eff}(Q)$ are shown as an inset.
The low-$Q$ increase with $S_\text{eff}(Q\rightarrow0)>1$ is due to enhanced inter-protein attractions upon approaching the LLPS (as well as minor changes of $c_\text{p}$).
A qualitatively similar behavior has been observed for other protein systems.\cite{Wolf2014,Moeller2014,Bucciarelli2016}

The model of Eq.~(\ref{eq:1}) is fitted to the experimental data where the analytical structure factor of adhesive hard-spheres in the Percus-Yevick approximation\cite{Menon1991,Menon1991b} is implemented and the form factor is described by a prolate ellipsoid of revolution\cite{Pedersen1997} with fixed parameters as determined in Section~\ref{sec:form}.
The experimental data are quantitatively reproduced by the model fits, in particular the low $Q$ upturn upon approaching phase separation.
For each scattering curve, the fit provides a refined value of the stickiness parameter $\tau$, which can be converted to a normalized second virial coefficient $b_2$ via Eq.~(\ref{eq:stick}).
The resulting $b_2$ values are displayed in Fig.~\ref{fig:saxs}(B) as a function of the reduced temperature $T/T_\text{c}$ of the solution.
The statistical uncertainty of $b_2$ is estimated to be $\pm 0.4$ based on the analysis of several independently prepared samples at the same condition.
In the temperature range investigated, $b_2$ increases monotonously with reduced temperature.
The interaction parameter retrieved from fitting, $\tau$, (and thus also $b_2$) as well as the quality of the fit are very similar if the form factor is modelled using the atomic coordinates implemented in a home-written programme that also takes the hydration layer into account\cite{Steiner2018}.

In addition to the $b_2$ data retrieved by the SAXS analysis, static light scattering (SLS) data\cite{Goegelein2012} on the aqueous system are shown.
Both data sets quantitatively agree with each other.
Thus, despite the proximity to LLPS, SAXS yields reliable results for the interaction parameters. 
Due to the quantitative agreement with the SLS data, the SAXS data thus provide further support for a universal temperature dependence (with respect to $T_\text{c}$) of $b_2$ of protein solutions, as previously noted for lysozyme\cite{Platten2015} and $\gamma$B-crystallin\cite{Bucciarelli2016}.

\subsection{Ethanol effect on the second virial coefficient of lysozyme solutions: Experiments and DLVO model}\label{sec:dlvo}

\begin{figure}
\includegraphics[width=\columnwidth]{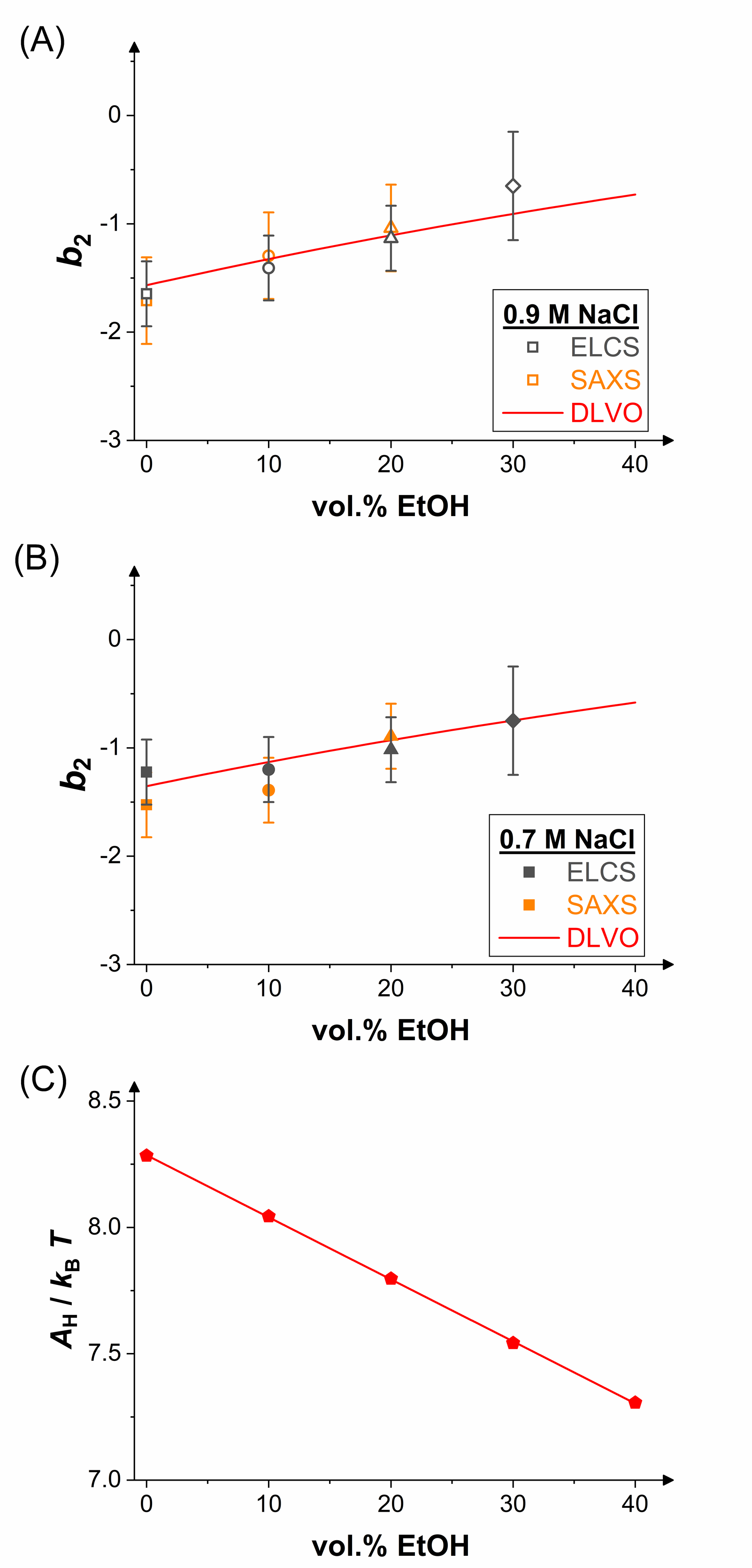} 
 \caption{Effect of ethanol on inter-protein interactions. (A) Normalized second virial coefficient $b_2$ as function of ethanol content at $25~^\circ\text{C}$ and in the presence of 0.9~M NaCl. 
 Data inferred from the cloud-point measurements (Fig.~\ref{fig:state}(A)) via the ELCS (grey symbols) and from the analysis of the SAXS data (Fig.~\ref{fig:saxs}(A)) (orange symbols) as well as calculated values based on the DLVO theory (line). (B) Data and theoretical predictions as in (A), but in the presence of 0.7~M NaCl. (C) Hamaker constant calculated based on Eq.~(\ref{eq:a}) (symbols) and linear fit (line). }
  \label{fig:b2et}
\end{figure}

Upon addition of ethanol, the LLPS coexistence curve of lysozyme shifts to lower temperatures, as discussed in Section~\ref{sec:llps}.
Accordingly, for increasing ethanol content, $T/T_\text{c}$ increases for a fixed temperature $T$, such as the temperature of the SAXS experiments (mainly $25~^\circ\text{C}$ but also a few at $20~^\circ\text{C}$).
The correspondingly reduced net attractions are reflected in the reduced low $Q$ scattering.
Hence, $b_2$ is expected to increase with ethanol content.
To quantify this dependence, values of $b_2$ were determined by comparing the low-concentration branches of the binodals (Fig.~\ref{fig:state}(A,B)) with those of SW fluids and exploiting the ELCS\cite{Platten2015}.
The results reveal a weak, but systematic increase of $b_2$ with ethanol content (Fig.~\ref{fig:b2et}(A,B)). 
Moreover, the values quantitatively agree with $b_2$ values determined by SAXS model fits (Figs.~\ref{fig:saxs}, \ref{fig:b2et}(A,B)).

In order to rationalize the dependence of $b_2$ on ethanol and salt content, protein interactions are modelled by the DLVO potential:\cite{Israelachvili1992}
\begin{eqnarray}
U_\text{DLVO}(r) = U_\text{HS}(r) + U_\text{SC}(r) + U_\text{VDW}(r) 
\end{eqnarray}
with the hard-sphere contribution $U_\text{HS}(r)$, the screened Coulomb contribution $U_\text{SC}(r)$ and the van der Waals contribution $U_\text{VDW}(r)$.
For $r > \sigma_\text{p}$ with particle diameter $\sigma_\text{p}$, $U_\text{SC}(r)$ is given by\cite{Poon2000}
\begin{eqnarray}
U_\text{SC}(r) = \frac{\left( Z e \right)^2}{4 \pi \epsilon_0 \epsilon_\text{s}} \frac{\exp{\left[ -\kappa (r - \sigma_\text{p}) \right]}}{\left( 1 + \kappa \sigma_\text{p}/2 \right)^2 r} 
\end{eqnarray}
with the number of positive elementary charges $Z=11.4$ at the present $p$H 4.5, the permittivity of the vacuum $\epsilon_0$ and that of the solvent $\epsilon_\text{s}$ as well as the Debye screening length $\kappa^{-1}$:
\begin{eqnarray}\label{eq:ka}
\kappa^2 =  \frac{e^2 N_\text{A}}{\epsilon_0 \epsilon_\text{s} k_\text{B} T} \sum_i{z_i^2 c_i} \,,
\end{eqnarray}
where $N_\text{A}$ is Avogadro's number and $z_i$ and $c_i$ are the valence and molar concentration of the $i$-th ionic species, respectively. 
The van der Waals component of the potential reads\cite{Poon2000} 
\begin{eqnarray}
U_\text{VDW}(r) = - \frac{A}{12} \left( \frac{\sigma_\text{p}^2}{r^2-\sigma_\text{p}^2} + \frac{\sigma_\text{p}^2}{r^2} + 2 \ln{\left[ 1 - \frac{\sigma_\text{p}^2}{r^2} \right]} \right) 
\end{eqnarray}
with the Hamaker constant $A$, which for two identical particles in a medium can be approximated by\cite{Israelachvili1992}
\begin{equation}\label{eq:a}
    A=\frac{3}{4} k_\text{B}\,T \left( \frac{\epsilon_\text{p}-\epsilon_\text{s}}{\epsilon_\text{p}+\epsilon_\text{s}} \right)^2 + \frac{3\,h\,\nu}{16\sqrt{2}} \frac{(n_\text{p}^2-n_\text{s}^2)^2}{(n_\text{p}^2+n_\text{s}^2)^{3/2}}
\end{equation}
with the permittivity of the protein $\epsilon_\text{p}$, the refractive indices of the particle, $n_\text{p}$, and that of the solvent $n_\text{s}$, Planck's constant $h$, and a characteristic ultraviolett absorption frequency $\nu$.
Optical and dielectric constants are reported in the literature both for the solvents (water--ethanol mixtures)\cite{Akerlof1932,Scott1946,Puranik1994,Herraez2006,El-Dossoki2007} and the particles (proteins)\cite{Farnum1999,Dwyer2000}.
For consistency with previous work,\cite{Sedgwick2007} we use $\sigma_\text{p}=3.4~\text{nm}$, $\epsilon_\text{p}=2$, $n_\text{p}=1.69$ and $\nu = 3\times 10^{15}~\text{s}^{-1}$.
With these values, Eq.~(\ref{eq:a}) has been applied to compute the Hamaker constant as a function of the ethanol concentration.
The result is shown in Fig.~\ref{fig:b2et}(C) as symbols.
Without ethanol, $A = 8.3~k_\text{B}T$ in agreement with previous studies.\cite{Poon2000,Sedgwick2007,Platten2015} 
Upon addition of ethanol, $A$ exhibits an apparently linear decrease (line),
reflecting lowered net inter-particle attractions in water--ethanol mixtures.
Similarly, a roughly linear decrease of $A$ was observed for water-glycerol and water-dimethyl sulfoxide mixtures.\cite{Goegelein2012,Platten2016}

The normalized second virial coefficient, $b_2$, can be computed based on the DLVO potential via Eq.~(\ref{eq:b}).
To avoid divergence of the integral, a cut-off length $\delta$ related to the Stern layer is used as lower integration limit. 
Its value, $\delta \approx 0.16~\text{nm}$, has previously been adjusted to match SLS data.\cite{Platten2016}
Note that $\sigma_\text{p} + \delta \approx \sigma$; i.e., the diameter of the adhesive hard sphere $\sigma$ assumed in the structure factor modelling agrees with that of the hard sphere amended with a cut-off layer used in the DLVO model.
This DLVO description might thus implicitly also account for non-DLVO effects, such as hydration, the hydrophobic effect and hydrogen bonding.\cite{Pellicane2004}
The results of the DLVO model are shown in Fig.~\ref{fig:b2et}(A,B) as solid lines.
The model $b_2$ monotonously increases with ethanol content and quantitatively agrees with the experimental data (symbols).
Thus, the ethanol-dependent changes of the inter-protein interactions are fully accounted for by its effect on the dielectric solution properties and thus on the Hamaker constant and the screening length.
Salt effects are contained in $\kappa$ (Eq.~\ref{eq:ka}).

Liu et al.\cite{Liu2004} studied lysozyme pair interactions in water--ethanol mixtures at neutral $p$H.
At three different NaCl concentrations, an increase of $b_2$ at low and a plateau at moderate ethanol concentrations was observed by light scattering.
To describe their data, they used a modified DLVO model.
The Hamaker constant $A$ was assumed to be constant irrespective of the ethanol content and the net charge $Z$ was treated as an adjustable parameter in contrast to our approach (Fig.~\ref{fig:b2et}(C)).
In addition, the DLVO potential was supplemented by an alcohol-dependent patchy SW potential and the interaction strength of the patch was allowed to vary with alcohol concentration.  
Compared with this more complex model, our DLVO calculation is simpler and does not require any free parameter.
However, if applied to their solution conditions, our model (with $Z=8.4$) predicts a monotonous weak increase of $b_2$ with ethanol content, similar to our experimental finding, and thus does not fully explain their observation.
It is conceivable that the slightly different trends observed are due to differences in accounting for the peculiar physicochemical properties of water--ethanol mixtures, such as the interpretation of $p$H values\cite{Bates1963}.

\section{Conclusion}

The phase behavior and the interactions of proteins in water--ethanol mixtures were studied.
The addition of moderate amounts of ethanol was found to decrease LLPS temperatures indicating reduced net attractions and, consistently, the low-$Q$ SAXS intensity of concentrated protein solutions decreases and the $b_2$ values increase.
The data suggest universal net interactions close to phase separation, supporting the extended law of corresponding states. 
The increase of $b_2$ with ethanol can be quantitatively captured by a DLVO model taking into account the effect of ethanol on the dielectric solution properties.
Thus, the DLVO theory can provide a mechanistic description of protein interactions also in complex solution environments.

\section*{Conflicts of interest}
There are no conflicts to declare.

\section*{Acknowledgements}
We thank Jan K.G. Dhont (FZ Jülich, Germany) and  Ram\'on Casta\~neda-Priego (Leon, Mexico) for stimulating and very helpful discussions, and Beatrice Plazzotta for assistance with the SAXS measurements in Aarhus.
F.P. acknowledges financial support by the Strategic Research Fund of the Heinrich Heine University (F 2016/1054-5) and the German Research Foundation (PL 869/2-1).
We thank the Center for Structural Studies (CSS) for access to the SAXS instrument.
CSS is funded by the Deutsche Forschungsgemeinschaft (DFG Grant numbers 417919780 and INST 208/761-1 FUGG).

%%%END OF MAIN TEXT%%%


\balance


%%%REFERENCES%%%
\bibliography{lorena} %You need to replace "rsc" on this line with the name of your .bib file
\bibliographystyle{rsc} %the RSC's .bst file
 
\newpage

\section*{Table of contents entry}
\begin{figure}[b!]
 \centering
\includegraphics[height=4cm]{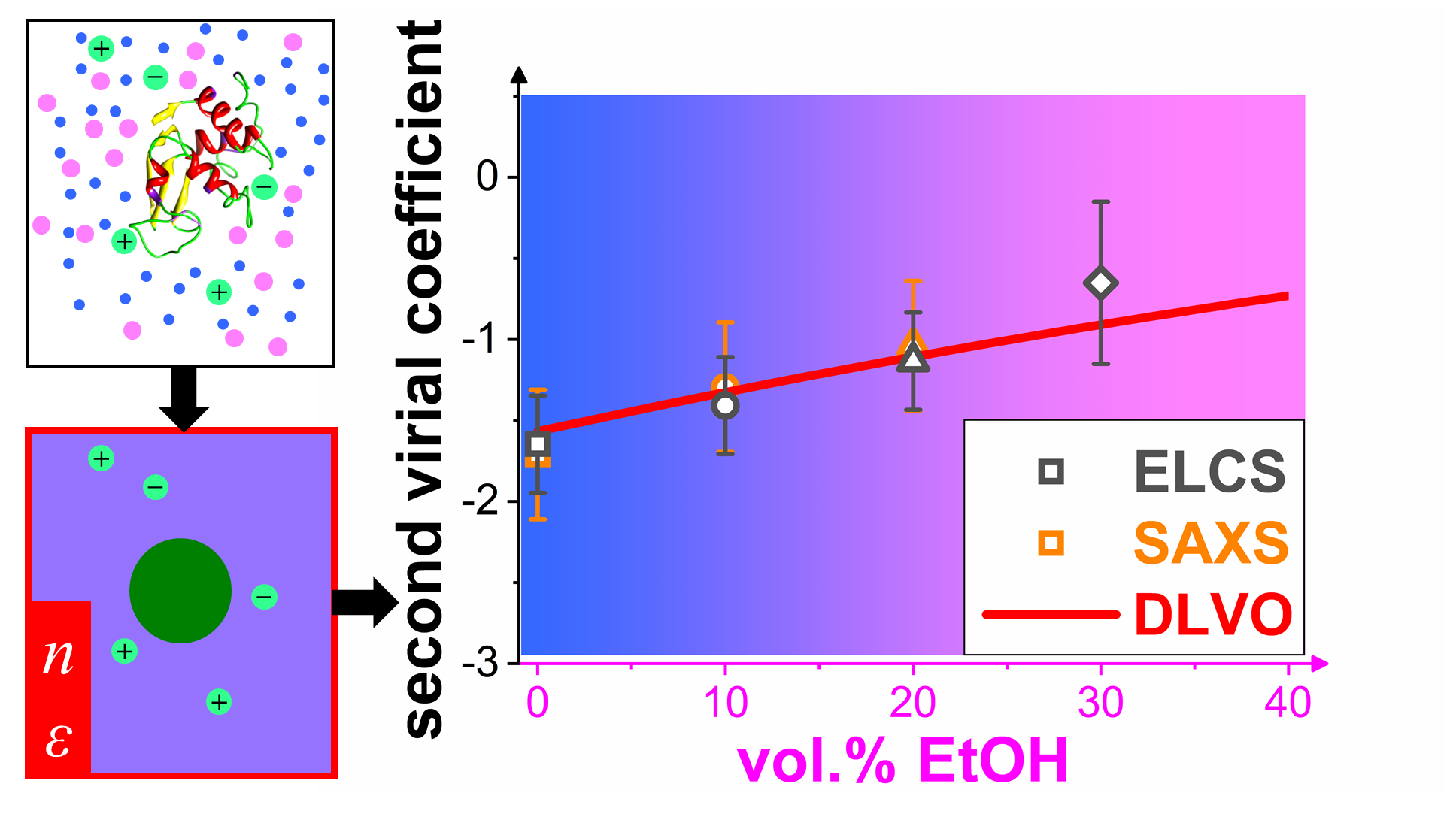}  
 \captionsetup{labelformat=empty}
\caption*{Adding a moderate amount of ethanol to a protein solution does not affect the protein shape and size but weakens the net inter-protein attraction. Corresponding changes of the liquid-liquid phase separation binodal and the SAXS intensity are observed. The effect of ethanol on protein interactions is entirely accounted for by its effect on the dielectric properties of the solvent.}
\end{figure}

\end{document}